\documentclass[12pt, a4paper]{amsart}

\usepackage{amsmath,amssymb,amsfonts}
\usepackage{bbold}
\usepackage{epic,eepic}
\usepackage{enumerate}

\usepackage{amsthm}
\usepackage[all]{xy}
\usepackage{caption}
\usepackage{stmaryrd} 
\usepackage[dvipdfmx]{graphicx,color} 
\usepackage{tikz}
\usetikzlibrary{trees}

\theoremstyle{plain}
\newtheorem{thm}{Theorem}

\theoremstyle{definition}

\numberwithin{figure}{section}

\numberwithin{table}{section}

\DeclareMathOperator*{\essinf}{ess\,inf}

\newcommand{\lspace} {
  \vspace{0.8\baselineskip}
}

\newdir{ >}{{}*!/-5pt/@{>}}

\definecolor{arrowred}{rgb}{0,0,0} 

\def \withName          {1}
\def \withAck           {0}

\begin{document}

\title[Optimality of VWAP Execution Strategies]{Optimality of VWAP Execution Strategies under General Shaped Market Impact Functions}

\if  \withName     1

\author[T. Kato]{Takashi Kato}
\email{kato@sigmath.es.osaka-u.ac.jp}
\fi  

\date{\today
}

\maketitle

\begin{abstract}
In this short note,
we study an optimization problem of expected implementation
shortfall (IS) cost under general shaped market impact functions.
In particular, 
we find that an optimal strategy is a VWAP
(volume weighted average price)
execution strategy
when the market model is
a Black-Scholes type 
with stochastic clock and market trading volume is large.
\end{abstract}

\lspace

Let
$
(\Omega, \mathcal{F}, 
(\mathcal{F}_t)_{t \ge 0},
\mathbb{P}
)
$
be a stochastic basis.
We consider
\begin{equation}
\label{eq:eq_1}
J(x)
  =
\sup_{
(\zeta_t)_{
    0 \le t \le T
  }
}
\mathbb{E}\Big[
  \int_0^T
\zeta_t
S_t
dt
\Big]
\end{equation}
subject to
\begin{equation*}
d S_t
  =
S_t(
  \mu d  V_t
+
  \sigma d \tilde{B}_{V_t}
-
  g(\zeta_t / v_t) dV_t),
\end{equation*}
where
$x \ge 0$,
$\mu < 0$,
$\sigma \ge 0$,
$
g :
[0, \infty)
  \to
[0, \infty)
$,
$
V_t
  =
\int_0^t v_r dr
$,
$
(v_t)_t
$
is an 
$(\mathcal{F}_t)_t$-adapted continuous positive process
with
$
\mathbb{E}[V_T]
  <
\infty
$,
$
(
  \tilde{B}_{\tilde{t}}
)_{\tilde{t} \ge 0}
$
is an 
$
(\tilde{\mathcal{F}}_{\tilde{t}})_{\tilde{t} \ge 0}
$-Brownian motion,
and
$
\tilde{\mathcal{F}}_{\tilde{t}}
  =
\mathcal{F}_{V^{-1}_{\tilde{t}}}
$,
$
(
\zeta_t
)_t
$
is an
$(\mathcal{F}_t)_t$-progressively measurable 
non-negative process
satisfying 
$
\int_0^T \zeta_t dt
  \le
x
\;
a.s.
$.

Here, 
$S_t$
(resp., 
$\zeta_t$
)
denotes
a security price
(resp.,
a trader's selling speed of the security)
at time $t$.
$g$
is regarded as a market impact function.

Note that
(\ref{eq:eq_1})
is equivalent with minimizing
expected IS cost problem
($
x S_0 - J(x)
$).

When
$
v_t
  \equiv
1
$
and
$g$
is quadratic,
(\ref{eq:eq_1})
is studied
in Section 5.2 of
\cite{kato_2014},
and we see that an optimal
strategy is a TWAP
(time weighted average price)
execution,
that is,
selling by constant speed,
when
$x$
is not so large.

This note presents a generalization of the
above result.
We assume the following conditions for $g$.
\begin{enumerate}
\item[(A1)]
$
g \in C^1((0, \infty)) \cap C([0, \infty)),
$

\item[(A2)]
$
g(0) = 0,
$

\item[(A3)]
$
h := g' \ge 0.
$
Moreover, there is a 
$
\zeta_0 \ge 0
$
such that
$h$
is non-increasing on
$(0, \zeta_0]$,
and
strictly increasing on
$
[\zeta_0, \infty)
$,

\item[(A4)]
$
\lim_{
  \zeta \to \infty
}
  h(\zeta)
=
  \infty
$.
\end{enumerate}
Note that when
$
\zeta_0 = 0
$,
$g$
is strictly convex
on
$
[0, \infty)
$.

Then, we have:
\begin{thm}
\label{thm:1}
Let
$
\nu \ge \zeta_0
$
be a unique solution to
$
\nu h(\nu)
 -
g(\nu)
  =
- \mu
$.
If
$
x \le \nu V_T
\; a.s.
$,
it holds that
$
J(x)
  =
\frac{S_0}{h(\nu)}
(1 - e^{-h(\nu)x})
$.
The corresponding optimal execution strategy is
given as
$
\hat{\zeta}_t
  =
\nu
v_t
\mathbb{1}_{
 \{
   v_t \le x / \nu
 \}
}
$.
\end{thm}

\lspace

Next
we treat more realistic situation where the
trader's execution itself increases the market trading volume,
that is,
\begin{equation}
\label{eq:eq_2}
\hat{J}(x)
  =
\sup_{
(\zeta_t)_{
    0 \le t \le T
  }
}
\mathbb{E}\Big[
  \int_0^T
  \zeta_t
  \hat{S}_t dt
\Big]
\end{equation}
subject to
\begin{align*}
&
d 
 \hat{S}_t
  =
\hat{S}_t \big(
  \mu d 
    \hat{V}_t
+
  \sigma d
    \tilde{B}_{\hat{V}_t}
-
  \hat{g}(\zeta_t / \hat{v}_t)
  d \hat{V}_t
\big),
   \quad
\hat{v}_t
  =
v_t + \zeta_t,
\\&
\hat{V}_t
  =
\int_0^t
 \hat{v}_r dr,
   \quad
\int_0^t
  \zeta_t dt
\le
  x
\; a.s. .
\end{align*}
Here,
$
\hat{g}
  :
[0, 1)
  \to
[0, \infty)
$
satisfies
(A1) - (A3)
replacing $\infty$
with
$1$
and the condition
$
\hat{h}(1-)
  =
\infty
$,
where
$
\hat{h}
  =
\hat{g}'
$.
\begin{thm}
\label{thm:2}
Let
$
\hat{\nu}
  \in
[\zeta_0, 1)
$
be a unique solution to
$
\hat{\nu}
\hat{h}(\hat{\nu})
  -
\hat{g}(\hat{\nu})
  =
- \mu
$,
and set
$
\nu
   = 
\hat{\nu} / (1 - \hat{\nu})
$.
If
$
x \le 
\nu V_T \; a.s.
$,
it holds that
$
\hat{J}(x)
  =
\frac{
  \hat{S}_0
}{
  \hat{h}(\hat{\nu})
}
(1 -
  e^{-
    \hat{h}(\hat{\nu}) x
  }
)
$.
The corresponding optimal execution strategy is given as
$
\hat{\zeta}_t
  =
\nu
v_t
\mathbb{1}_{
 \{
    v_t \le x / \nu  
 \}
}
$.
\end{thm}

Proofs of
Theorem \ref{thm:1}
and
\ref{thm:2}
are not difficult.
We can show them
by standard
verification arguments.
In any case,
the optimality of VWAP execution strategies
is guaranteed
under general
$g$
(, $\hat{g}$)
whenever
$x$
is not so large
(in other words,
$
\essinf
V_T
$
is not so small).

\if \withAck   1
\section*{Acknowledgement}

The author thanks Prof. Takanori Adachi
(Ritsumeikan Univ.) for his dedicated support.
\fi 

\providecommand{\bysame}{\leavevmode\hbox to3em{\hrulefill}\thinspace}
\providecommand{\MR}{\relax\ifhmode\unskip\space\fi MR }
\providecommand{\MRhref}[2]{%
  \href{http://www.ams.org/mathscinet-getitem?mr=#1}{#2}
}
\providecommand{\href}[2]{#2}


\begin{thebibliography}{Mac97}

\bibitem[Kato14]{kato_2014}
Takashi Kato, \emph{An optimal execution problem with market impact}, Finance and
  Stochastics \textbf{18} (2014), no.~3, 695--732.

\end{thebibliography}
\end{document}